\documentclass[12pt]{article}
\usepackage{lineno}
\usepackage{scicite}
\usepackage{times}
\usepackage{graphicx}
\usepackage{color}

\newcommand{\figref}[1]{Fig.~\ref{#1}}

\newcounter{lastnote}

\title{Curved crack paths are predicted by elastic-charges}

\author
{Oran Szachter$^{1\dagger}$, Emmanuel Siéfert$^{2\dagger}$,  Mokhtar Adda-Bedia$^{3}$, \\ Eran Sharon $^{1}$, Michael Moshe $^{1\ast}$\\
\\
\normalsize{$^{1}$The Racah Institute of Physics, The Hebrew University of Jerusalem, Jerusalem, 91904, Israel}\\
\normalsize{$^{2}$Université Grenoble Alpes, CNRS, LIPhy, F-38400 Grenoble, France}\\
\normalsize{$^{3}$Laboratoire de Physique, CNRS, ENS de Lyon, Universit\'e de Lyon, F-69342 Lyon, France}\\
\\
\normalsize{$^\ast$To whom correspondence should be addressed; E-mail:  michael.moshe@mail.huji.ac.il}
\\
\normalsize{$^\dagger$These authors contributed equally to this work.}
}

\date{}

\begin{document} 

\baselineskip24pt

\maketitle

\begin{abstract}
{Predicting crack trajectories in brittle solids remains an open challenge in fracture mechanics due to the non-local nature of crack propagation and the way cracks modify their surrounding medium. Here, we develop a framework for analytically predicting crack trajectories, similar to predicting the motion of charged particles in external fields within Newtonian mechanics. We demonstrate that a crack can be described as a distribution of elastic charges, and within the framework of Linear Elastic Fracture Mechanics (LEFM), its interaction with the background stress can be approximated by a singular geometric charge at the crack tip. The crack’s motion is then predicted as the propagation of this singular charge within the unperturbed stress field. We apply our approach to study crack trajectories near defects and validate it through experiments on flat elastomer sheets containing an edge dislocation. The experimental results show excellent agreement with theoretical predictions, including the convergence of curved crack trajectories toward a single focal point. We discuss future extension of our theory to the motion of multiple interacting cracks. Our findings highlight the potential of the elastic-charges approach to significantly advance classical fracture mechanics by enabling analytical solutions to problems traditionally requiring numerical methods.}
\end{abstract}

Cracks are ubiquitous in nature, spanning a wide range of scales from microscopic~\cite{suo1992fracture, yao2008macro} to planetary~\cite{walwer2021magma}. They manifest in both natural and man-made materials, making them relevant in fields as diverse as engineering science~\cite{bavzant1984size,jenq1985two}, condensed matter physics~\cite{zhu2008dynamic,rafiee2010fracture,zhang2014fracture,shim2018controlled}, and biology~\cite{ohno1996progression,barthelat2007mechanics,ikuno2008lacquer}. 
Nucleation and propagation of cracks, as mechanical processes, hinge on elastic background stress fields generated by various sources such as external loading~\cite{broberg1999cracks}, spatially nonuniform fields~\cite{freund1972motion}, or topological defects~\cite{erdogan1974interaction}. Another example is that of tearing thin sheets, where propagation is dominated by bending moments \cite{ibarra2019predicting}.
Intriguingly in all these situations, crack propagation can occur along counter-intuitive curved trajectories. 
For instance, when two cracks approach each other, they may exhibit a pattern of repulsion followed by attraction, eventually closing a loop~\cite{fender2010universal,schwaab2018interacting}. This crack tip propagation bears a striking resemblance to the motion of interacting particles in Newtonian mechanics, where solutions for the single-body and two-body problems are well-established. Unlike the case of tearing, where the absence of stretching allows the prediction of curved trajectories, the prediction of stretching driven crack trajectories remains a central open problem in fracture mechanics, even for the simpler scenario of a single crack propagating in a residually stressed solid~\cite{adda1995crack,corson2009thermal}.

The current theoretical framework for crack propagation primarily relies on Linear Elastic Fracture Mechanics (LEFM), which combines the elastic problem defined by the crack edges and material geometry with criteria for crack propagation.  
The most popular criterion for crack propagation consists of two parts: the principle of local symmetry, which selects the crack advance along the direction with maximum tension and vanishing shear, and the Griffith energy balance, according to which propagation occurs only if released energy exceeds a threshold characterizing the specific material ~\cite{leblond2003mecanique,cotterell1980slightly}. In cases of quasi-static propagation, the crack remains in equilibrium at each step, enabling the calculation of elastic fields that promote crack growth and the determination of infinitesimal crack extension \cite{mitchell2017fracture, freund1972motion}. The same procedure must be repeated for the updated stress field after each infinitesimal advancement to predict crack propagation further. Since the crack surface itself extends during the process, solving the elastic problem as a boundary problem requires an iterative approach.
Consequently, their history can {significantly modify the stressed state, and thus} influence crack trajectories, distinguishing them from the Newtonian-like mechanics of particles. To avoid such an iterative procedure, variational quasistatic crack evolution models have been proposed in which the crack evolution is reduced to a global energy minimization problem. This approach paved the way for phase-field models ~\cite{francfort1998revisiting,bourdin2000numerical,bourdin2008variational}, which have established themselves as a robust numerical framework for studying crack propagation. 

{In this work, we approach the quasi-static crack propagation problem from a geometric perspective by reformulating the elastic problem and the laws of crack growth using elastic charges. We assume a separation of time scales, where the elastic wave speed far exceeds the crack propagation speed, allowing analysis based on mechanical equilibrium \cite{freund1998dynamic}. We show that the mechanical effect of a static crack in an elastic medium is equivalent to a distribution of elastic charges along the crack tail. Within the framework of LEFM, the elastic charges approach introduces a new approximation scheme, wherein the interaction of a static crack with the background stress is dominated by effective singular charges located at the crack tip. These singular charges, a monopole and a dipole, encode the force and torque acting on the crack through their interaction with the background stress, generalizing the well-known concept of the Peach-Koehler force to what we refer to as a “Peach-Koehler torque.”
We then reformulate the crack propagation criteria, which, in terms of elastic charges, reduce to equations of motion for the singular charges. As a result, the original problem is simplified to the quasi-static motion of singular charges, representing the extended crack and determining both its propagation and orientation at every point. This theoretical step enables the prediction of the entire crack trajectory all at once, analogous to the motion of a Newtonian particle in an external field, and allows for analytical solutions to the problem.
To test the analytical power of our framework, we examine the theoretical problem of a crack propagating in a large medium near a defect, a setup that features non-uniform stresses, which promote curved crack trajectories. We solve this class of problems, which is relevant to a broad range of physical systems, and derive analytical solutions for crack trajectories in closed form.
Finally, to validate our theory and its predictive power, we conduct experiments on the quasi-static propagation of cracks in a flattened dislocated annulus. Our framework predicts that cracks will follow curved trajectories that converge toward a focal point, with the location of this point depending on the system’s dimensions. Notably, our experimental results quantitatively confirm this prediction.}
Moreover, we find that when combined with the Griffith energy criterion, our theoretical approach predicts that crack propagation stops slightly before reaching the focal point, in agreement with experimental observations as well.
Overall, our geometric perspective, which characterizes cracks as distributed charges interacting with elastic fields, offers a comprehensive understanding of crack mechanics. This approach introduces a new paradigm for predicting crack trajectories and their interactions with defects. The experimental and theoretical agreement underscores its effectiveness and supports its potential application in various crack-related phenomena. 

{In the next section, we present the theoretical background, focusing on the reformulation of the quasi-static crack problem in terms of elastic charges. We then focus on the case of a single crack propagating in a non-uniformly stressed solid, and develop an analytical framework for predicting crack trajectories. This is followed by a description of the problem of a crack propagating in a defective medium, and the experimental setup designed to test the theory. We derive the corresponding theoretical predictions and compare them with experimental results. Finally, we conclude by discussing the broader implications of our work for addressing other challenges in fracture mechanics and outline potential future directions.}

\section*{Theoretical background}
\begin{figure*}
	\centering
	\includegraphics[width=0.8\linewidth]{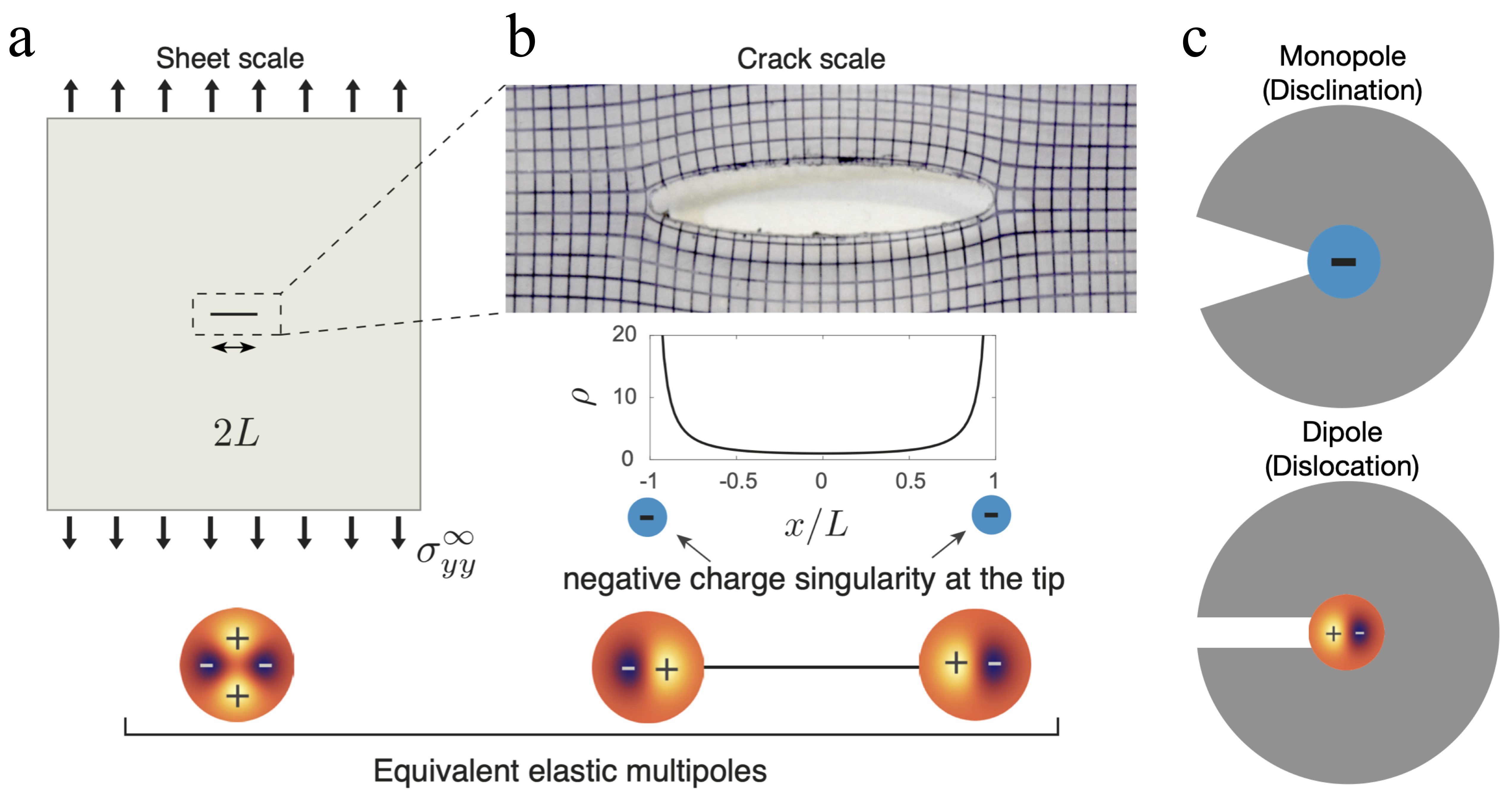}
	\caption{\textbf{Cracks as elastic multipole charges.} \textbf{a}. Within a  far field approximation, a small crack behaves as a quadrupole. \textbf{b.} At scales comparable to crack length, as well as at large distances, the crack is better approximated by two opposite dipoles sitting at the crack tip. \textbf{c.} The singular elastic charges induced by the angular (top) and linear (bottom) discontinuities induced by the crack opening.}
\label{fig:Figure1}
\end{figure*}

A crack is defined as a curve $\mathcal{C}$ along which the elastic fields can exhibit discontinuities. For instance, when the material is subjected to remote stresses and the crack surface is traction-free, {the displacement field experiences a jump across the crack. In mechanical equilibrium, the problem reduces to a divergence-free stress field subject to boundary conditions that require vanishing normal stress along the crack surface and prescribed stresses on the remote boundaries}
\begin{equation}
	\mathrm{div} \sigma  = 0 \;,\quad 	\sigma \cdot  \mathbf{n} \left.\right\vert_{\mathcal{C}} = 0\;, \quad 	\sigma \cdot  \mathbf{n} \left.\right\vert_{{\infty}} = \sigma_\mathrm{ext}\;.
	\label{eq:Equilibrium}
\end{equation}
{A key step in this work is the reformulation of the static crack problem using elastic charges, which have been established as a geometric description of singular sources of stress \cite{moshe2015geometry}. The idea is most intuitively understood through an electrostatic analogy, where a traction-free crack is analogous to a conductive needle, and the background stress corresponds to an external electric field. In this analogy, two equivalent mathematical formulations arise: (1) a boundary value problem for the electric field, satisfying the static Maxwell equations with the appropriate boundary conditions on the needle, and (2) an optimization problem for the electric charge distribution along the needle, which minimizes the total energy stored in the electric field \cite{griffiths1996charge,jackson2000charge}.

Based on this analogy we reformulate the static crack problem. A distribution of elastic charges $\rho$ along a curve $\mathcal{C}$ induces a stress field 
\begin{equation}
	\sigma^{\mu\nu}(\mathbf{x}) = \int_{\mathcal{C}} G^{\mu\nu}(\mathbf{x} - \mathbf{x}') \rho(\mathbf{x}') \mathbf{d} \mathbf{x}' \;,
	\label{eq:Green}
\end{equation}
where $G^{\mu\nu}$ represents the stress induced by a single (monopole) elastic charge, and $\mathbf{x}'$ is the coordinate along the crack curve $\mathcal{C}$. Minimizing the total elastic energy stored in the stress field with respect to the charge distribution $\rho$ is equivalent to solving the elastic equilibrium problem described in Eq.~\ref{eq:Equilibrium}. These induced charges form to facilitate the stress-free boundary conditions on the crack faces.
Therefore the boundary value problem of a static crack is mathematically equivalent to finding the energy minimzing charge distribution along the curve describing the crack. The mathematical details of this equivalence are provided in the appendix for both the electric and elastic cases (see “Equivalence of charge and boundary conditions” in the supplementary materials). 

At the core of the crack propagation problem is the interaction between the crack and the background stress, which can be understood as the interaction between the charge distribution along the crack and the background stress sources. While the charge-based formulation offers general advantages for numerical implementations, it does not inherently simplify the problem from an analytical perspective. However, within the framework of LEFM, we will show that this problem simplifies significantly, and predicting curved crack trajectories requires only the key characteristics of the charge distribution rather than its full detailed calculation. Therefore, for the remainder of this paper, we focus on implementing our method within the LEFM framework, specifically using the formalism of a stress potential, commonly known as Airy’s stress function \cite{landau1959course}. Within linear elasticity, the stress potential satisfies the equation 
\begin{equation}
	\frac{1}{Y}\Delta \Delta \psi = K\;,
	\label{eq:chargeSF}
\end{equation}
where $K = K_0 + K_\mathcal{C}$ is the distribution of elastic charges representing the background and crack sources} 
and the stress, $\sigma$, is derived from the stress function $\sigma^{\alpha\beta} = \epsilon^{\alpha\mu} \epsilon^{\beta\nu} \partial_{\mu\nu} \psi$.
{If $x_{1}$ and $x_2$ are coordinates along the directions parallel and orthogonal to crack curve $\mathcal{C}$, then $K_\mathcal{C} =  \rho({x_1}) \delta({x_2})$. 
From linearity, the solution of this equation is written in the form
\begin{equation}
    \psi = \psi_0 + \psi_\mathcal{C}\;,
    \label{eq:SFdecomp}
\end{equation}
where, $\psi_0$ describes the background stress, and $\psi_\mathcal{C}$ represent the stresses induced by the crack, expressed as a formal solution in terms of the unknown charge density $\rho$, similar to Eq.\ref{eq:Green}.}
Before continuing to formulate the crack growth problem, we demonstrate the concept of elastic charges in the simplest problem of a finite crack of length $2L$ in an infinite medium subjected to remote uni-axial stress $\sigma_{yy}^{\infty}$ along the $y$ direction, as illustrated in \figref{fig:Figure1}A. In this case, we have 
\begin{eqnarray}
	\psi_0(\mathbf{x}) = \frac{1}{2} \sigma_{yy}^{\infty} x^2 \;,\quad 
	\psi_\mathcal{C}(\mathbf{x}) = \int G(\mathbf{x} - \mathbf{x}')K_\mathcal{C}(\mathbf{x}') \mathrm{d} \mathbf{x}'\;.
	\label{eq:finite}
\end{eqnarray}
where $G$ is the Green’s function for an infinite domain with vanishing stress at infinity \cite{moshe2015elastic}. The charge distribution that satisfies the stress-free boundary conditions and minimizes the total energy is given by (\figref{fig:Figure1}B, see "Charge Density" in the supplementary materials):
\begin{equation}
	K_\mathcal{C} (\mathbf{x} )= \rho(x) \delta(y) \mathrm{\quad with \quad} \rho(x) = C \left[1 - \left(\frac{x}{L}\right)^2\right]^{-3/2}.
	\label{eq:FiniteCrackCharge}
\end{equation}
{This charge density is supplemented with two negative singular charges at crack ends to comply with the conserved vanishing monopole. Notably, it exhibits a power-law divergence $\rho \sim r^{-3/2}$, reflecting the well-known universal square-root singularity of stress at the crack tip. This result implies the universality of the singular charge distribution (see "Asymptotic Behavior" in the supplementary materials).}  

Returning to the general case of a curved quasi-static crack $\mathcal{C}$ with an energy-minimizing charge density $\rho$ at a given instant, its subsequent propagation is governed by two complementary criteria. First, the principle of local symmetry states that the crack propagates in the direction of maximal tensile stress, which corresponds to vanishing torque on the crack~\cite{gurtin1996configurational}, ensuring mode-I propagation. Second, the Griffith criterion determines whether the crack will advance: propagation occurs only if the energy released per unit crack extension exceeds the Griffith energy, a material-dependent property. Both criteria depend on the interaction between the crack and the background stress.

One advantage of the charges approach is that it provides a straightforward recipe for calculating interactions between different sources of stresses, {including curved cracks}. 
{The potential energy stored in the elastic fields is \cite{moshe2015elastic}}
\begin{equation}
	U = \int K \,\psi \,\mathrm{d}^2 \mathbf{x}\;.
	\label{eq:int}
\end{equation}

Consequently, the interaction energy contains a constant $U_0$ representing the self-energy $(\sim K_0\psi_0)$ of the background stress and two additional terms: The self-interaction energy $(\sim K_\mathcal{C}\psi_\mathcal{C})$ of the induced charges along the crack, and their interaction $(\sim K_\mathcal{C}\psi_0+K_0\psi_\mathcal{C})$ with the background stresses. 
While the latter can be either positive or negative, the former is positive-definite, restricting the emergent charge distribution $K_\mathcal{C}$.
Note that Eq.~\ref{eq:int} requires modifications when finite domains are under consideration, especially in the form of boundary terms (see "Interactions in finite domains" in the supplementary materials). 
Notably, the total charge distribution along the crack should be considered to calculate the interaction energy. 

{We can now reformulate the crack growth problem as follows:\\
\textit{Consider an elastic medium  $\Omega $ with background stress sources  $K_0$  and a crack  $\mathcal{C}$ . Determine the energy-minimizing charge distribution  $\rho$  along $\mathcal{C}$  and compute its interaction with  $K_0$. Upon infinitesimal crack extension, find the energy-minimizing propagation direction and evaluate the corresponding change in potential energy. If the released energy exceeds Griffith’s threshold, advance the crack accordingly and iterate the process.}

This formulation, while providing a new perspective, does not inherently simplify the problem compared to standard approaches in LEFM. However, in the next section, we show that under the assumption of a static  $K_0$  and a crack sufficiently far from other stress sources, this seemingly complex task simplifies significantly.}

\section*{Elastic charges theory for crack path prediction}
In this section, we examine crack propagation in the special case where the driving stresses originate from a remote, static source  $K_0$ . While the local stress state near the crack may be complex, its interaction energy  with  $K_0$  can be fully described by a multipole expansion \cite{JacksonJohnDavid1999Ce}. The first step in this formulation is identifying the center of charge relative to which this expansion is performed. A key insight comes from the case of a finite crack  (\figref{fig:Figure1}).

At distances much larger than the crack length $( L \ll r )$, the net monopole and dipole charges vanish, making the quadrupole moment the leading-order contribution, centered at the middle of the crack. 
However, the power-law divergence of the charge distribution implies that the center of charge for each half of the crack is located precisely at the crack tip (see "Center of Charge" in the supplementary materials for a detailed derivation). Consequently, the total quadrupole moment is better approximated as a pair of opposite dipoles positioned at the crack tips (\figref{fig:Figure1}B). More generally, the universal square-root singularity in LEFM~\cite{williams1952stress,williams1957stress} thus translates into a universal power-law singularity of the charge distribution, ensuring that, regardless of its detailed structure, the effective center of charge always resides at the crack tip. Explicitly, upon assuming that the crack tip is located at $\mathbf{x}_c$, the approximated charge density is
\begin{eqnarray}
	K_\mathcal{C} = m \, \delta (\mathbf{x} - \mathbf{x}_c) + \mathbf{p} \cdot \nabla \delta (\mathbf{x} - \mathbf{x}_c) + \dots\;,
	\label{eq:CrackMultipole}
\end{eqnarray}
with $m$ the monopole moment, $\mathbf{p}$ the dipole moment, etc.
Notably, monopole and dipole moments induce long-range fields, whereas higher-order multipoles are short-ranged~\cite{moshe2015elastic}. As a result, the interaction energy between the crack and a remote source of stress is dominated by the two singular charges at the crack tip, although the stressed state of a crack can be complex.
This approximation is expected to hold in situations where the singular part of the stress field induced by the crack dominates the interaction, that is, when sources of stresses are remote.
Additionally, as illustrated in \figref{fig:Figure1}C, these monopole and dipole charges represent localized objects that encode extended discontinuous deformations, explaining how singular charges effectively describe the interaction of an extended object like a crack. 
Thus, crack propagation is equivalent to the propagation of a point-like object containing two multipolar charges, just as in Newtonian mechanics of particles. 
The interaction energy of a crack located at $\mathbf{x}_c$ with the background stress, which would be crucial for predicting crack trajectories, then takes the form
\begin{equation}
	U = \int \psi_0(\mathbf{x})\, \left[ m \delta(\mathbf{x} - \mathbf{x}_c) + \mathbf{p} \cdot \nabla \delta(\mathbf{x} - \mathbf{x}_c)\right] \mathrm{d}^2 \mathbf{x} \;. 
	\label{eq:InfPotential}
\end{equation} 

Next, we reformulate the propagation criteria described in the previous section within the singular charge approximation. Beginning with the principle of local symmetry, which determines the local orientation of the crack, we note that the torque acting on the crack is equivalent to the torque acting on the singular dipole moment $\mathbf{p}$. Consequently, this criterion reduces to propagation along the direction of $\mathbf{p}$, ensuring that the torque on it vanishes. By differentiating the potential energy with respect to the dipole orientation, we obtain the torque:
\begin{equation}
	\tau = \ p_\beta \epsilon^{\alpha\beta} \nabla_{\alpha} \psi_0
	\label{eq:torque}
\end{equation}
This expression generalizes the classical Peach-Koehler force, which governs the motion of dislocations by considering forces on a dipole moment, to what we term a Peach-Koehler torque, which governs the crack’s orientation via the torque acting on its singular dipole moment.
In classical elasticity, the rotation of $\mathbf{p}$ is prohibited due to the topological conservation of the dipole moment. However, in our formulation, this restriction is lifted because the dipole is located at a boundary, where it is locally non-conserved.
The crack orientation is determined by the condition $\tau = 0$, which implies that $\mathbf{p}$ aligns with the gradient of $\psi_0$. This allows us to construct a vector field representing the directions of $\mathbf{p}$ at every point. Since a crack propagates along $\mathbf{p}$, its trajectory follows the streamlines of this field, providing a natural framework for predicting crack growth. Note that predicting crack trajectories as streamlines of a field determined solely by the background stress represents a major advancement over standard approaches in LEFM, which typically require iterative solutions to the crack growth problem.

We now turn to reformulating the Griffith criterion, which states that crack propagation occurs only if the energy released by the advancing crack exceeds the fracture energy required for material separation. Within the charges formulation and the singular charge approximation, this criterion translates to the energy released by the propagation of the monopole and dipole upon an infinitesimal crack extension. This is directly analogous to the force acting on the crack tip, where the Griffith threshold can be interpreted as a static friction force resisting propagation. The force acting on the crack, derived from the potential energy, is given by:
\begin{equation}
	\mathbf{F} = \left(m  - \mathbf{p}\cdot \nabla \right)\nabla \psi_0
	\label{eq:force}
\end{equation}
First, we note that while crack trajectory predictions rely solely on the direction of $\mathbf{p}$, the energy release rate, and thus the force driving crack propagation, explicitly depends on the magnitudes of singular charge multipoles $m$ and $\mathbf{p}$. These can be determined by minimizing the potential energy with respect to $m$ and $\mathbf{p}$, rather than solving for the full charge density $\rho$. Second, in general, the force acting on the crack tip is dominated by the monopole, and the dipole term decays faster with the distance of the crack from the source of $\psi_0$. This is indicated by the extra derivative, thus we can approximate $\mathbf{F}  \approx m \nabla \psi_0$. However, finite cracks are constrained to have vanishing monopole and dipole, thus $m = 0$. In this case $\mathbf{F} = \left( \mathbf{p}\cdot \nabla \right)\nabla \psi_0$. In either case, the parametric curve describing crack trajectory satisfies $\dot{\gamma}(s) =  \mathbf{F} - \mathbf{F}_G$ with $F_G$ the Griffith energy release rate. Note that $m$ or $\mathbf{p}$ can still be position dependent, thus their computation is necessary. However, the construction of crack trajectory depends on $\nabla \psi_0$ only.
The theoretical framework developed here provides a systematic approach for analytically computing crack trajectories and propagation conditions. The key steps are as follows:\\
\textit{Consider an elastic medium $\Omega$ with background stress sources $K_0$ described by the stress function $\psi_0$, and a crack $\mathcal{C}$. For every point in $\Omega$, determine the energy-minimizing values of $m$ and $\mathbf{p}$ by substituting Eq.\ref{eq:CrackMultipole} into Eq.\ref{eq:int}.
Compute the torque-free orientation of $\mathbf{p}$ using Eq.\ref{eq:torque}, and the force $\mathbf{F}$ from Eq.\ref{eq:force}.
The crack trajectories are given by the streamlines of $\mathbf{p}$, and actual propagation occurs if $\mathbf{F}$ exceeds the Griffith energy release rate.}

Before exploring setups with curved crack trajectories, we highlight key implications of this formulation.
First, a remarkable result relevant for finite crack, is that due to the conservation of net dipole \cite{kupferman2015metric}, two crack ends are mechanically entangled: Reorientation of one tip, which corresponds to the reorientation of the dipole moment at one side, necessitates the opposite reorientation of the other tip. Thus, if one tip is held fixed, reorientation of the other crack tip is topologically prohibited. 
Second, we find that curved crack paths depend solely on the pre-stressed state $\psi_0$, meaning the director field of energy-minimizing dipoles can be determined without solving for the full charge distribution $\rho$ or its singular multipoles $m$ and $\mathbf{p}$. While the stress state of the system is influenced by $\rho$, the interaction of the crack with the background, and thus its trajectory, is fully determined by $\psi_0$.
Third, crack trajectories depend only on the crack tip’s initial position, not on the detailed history of the initial crack $\mathcal{C}$. This contrasts with the conventional view that crack propagation is history-dependent. We will test this prediction experimentally in a later section. However, we stress that this result holds only for a single crack interacting with static stress sources. It breaks down in cases where multiple cracks interact, a scenario discussed further in the summary section.

\begin{figure}
	\centering
	\includegraphics[width=0.4\linewidth]{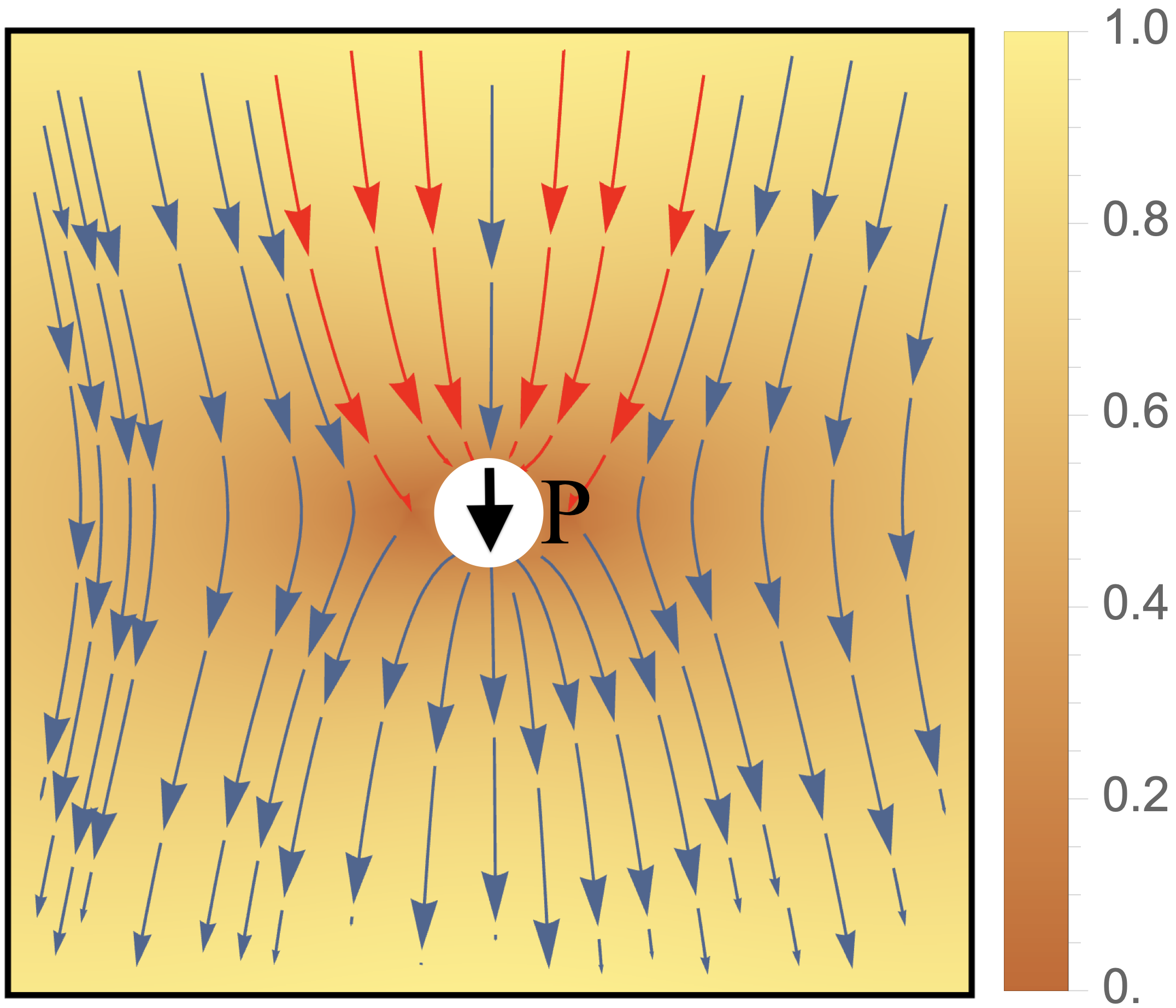}
	\caption{Predicted crack trajectories at the vicinity of a dislocation described by a dipole $\mathbf{P}$ (the black arrow). The trajectories are streamlines of the gradient of the stress function in Eq.\ref{eq:dipsf}. The red arrows indicate a convergence effect due to the background dislocation. The colored background represents the (rescaled) energy release upon assuming that the force acting on the crack is dominated by the dipole. }
	\label{fig:Figure2}
\end{figure}

With these predictions, we now examine crack trajectories near defects in a large solid. Recent experiments have shown that defects can influence crack motion \cite{rozen2020fast}. To explore this effect, we consider the propagation of a crack in a large medium containing a dislocation, an essential problem in materials science. In this problem the source of stress is static, and we expect our theory to hold as long as the crack is not too close to the defect. 
The stress function associated with a background dislocation, characterized by a dipole $\mathbf{P}$ (as apposed to $\mathbf{p}$ which characterized the crack's dipole), is given by \cite{seung1988defects, moshe2015elastic}:
\begin{equation}
	\psi_0^\mathrm{dis}  = \frac{Y}{8\pi} \mathbf{P} \cdot \mathbf{x} \ln |\mathbf{x}|^2 ;.
	\label{eq:dipsf}
\end{equation}
According to Eq.~\ref{eq:torque}, crack trajectories follow the streamlines of the vector field $\nabla \psi_0^\mathrm{dis}$, which leads to a differential equation for the trajectories. This equation admits a closed-form solution,
\begin{eqnarray}
	x(y) = \sqrt{P^2 e^{-2 \alpha P^2 y} - y^2},
	\label{eq:DipoleTraj}
\end{eqnarray}
where $\alpha$ is an integration constant dependent on the initial position. The streamlines are shown in \figref{fig:Figure2}, where the background heat map represents the magnitude of $\nabla \psi_0^\mathrm{dis}$, which is proportional to the energy release rate. This suggests that cracks will propagate from top toward the dislocation as illustrated by the red streamlines in \figref{fig:Figure2}, but may halt before reaching it due to the decreasing energy release rate. Notably, cracks originating from the bottom do not propagate, as they must advance in the direction of the dipole field. While cracks can propagate from the dislocation outside downwards, such cracks would modify the source charge $K_0$, and requires treatment that is beyond the scope of this work. 
The analytic calculation of cracks trajectories at the vicinity of defects, can be repeated for any singular source of stress. In the next section we turn to test our theoretical predictions experimentally. 

\section*{Crack trajectories in a dislocated annular domain}
\begin{figure*}
	\centering
	\includegraphics[width=0.9\linewidth]{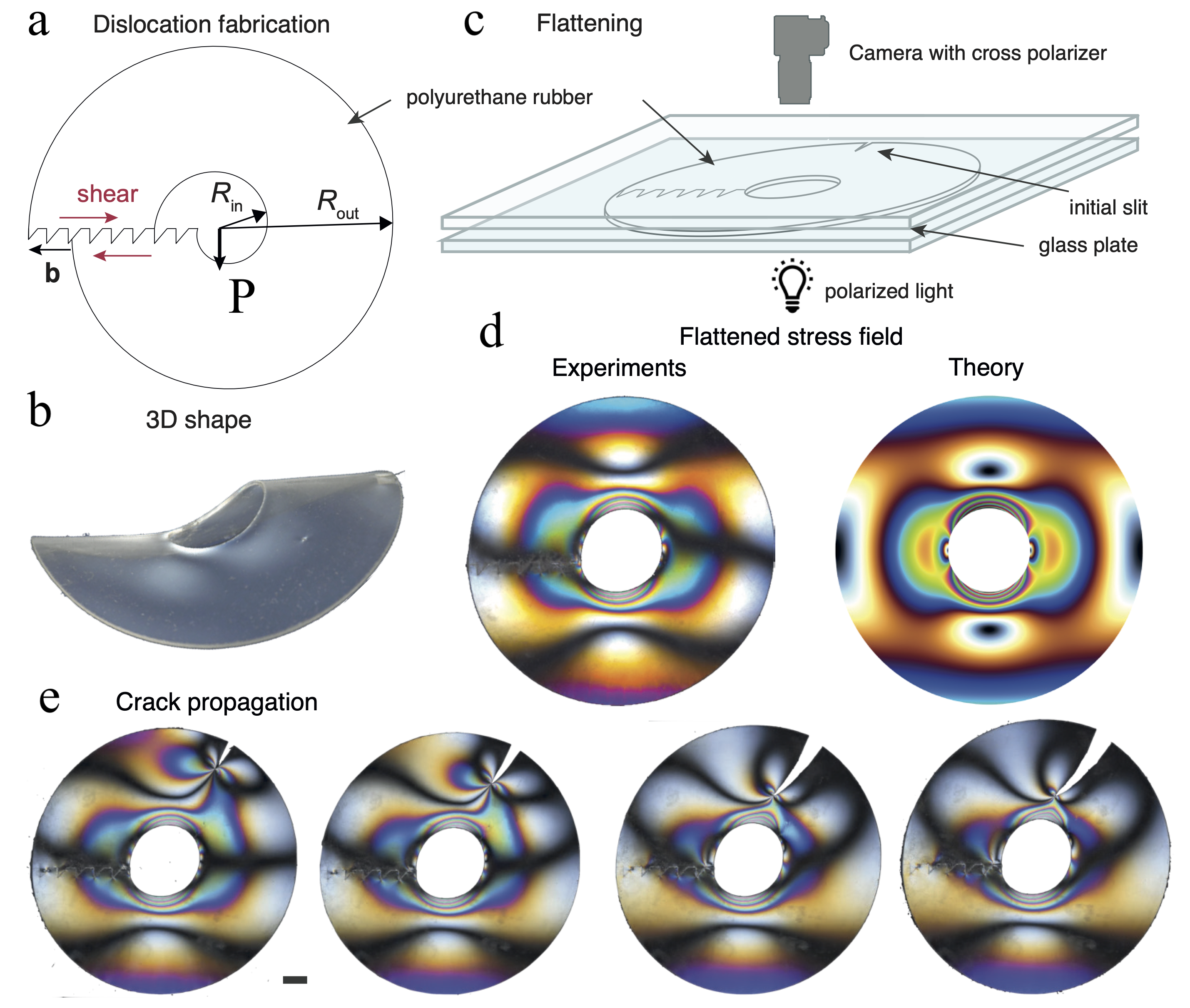}
	\caption{\textbf{Cracks around a dislocation.} \textbf{a}. Fabrication of a dislocated annulus made of a photoelastic polyurethane rubber, with $R_\mathrm{in}$ the internal and $R_\mathrm{out}$ the external radii, $\bf{b}$ the Burger's vector and $\mathbf{P}$ the resulting dipole source term. The annulus is sheared along a jigsaw pattern and then glued using the same uncured material as glue. \textbf{b.} Resulting 3D shape. \textbf{c.} The structure is then coated with talcum powder, flattened between two glass plates, and observed through cross polarizers to see the stress field. \textbf{d.} In the absence of cracks, the residual stress field is well described by linear elasticity~\cite{seung1988defects}. \textbf{e.} A crack is initiated at the boundary, propagates through the structure along a curved trajectory, and eventually stops. The complete propagation typically takes one hour. Scale bar: 1cm. }
	\label{fig:Figure3}
\end{figure*}
To experimentally test our approach, we study a finite domain containing a dislocation. Specifically, we consider an annular dislocated domain, a geometry that introduces boundary conditions that necessitate modifications to the predictions derived in the previous section.
We manufacture thin annular sheets with inner and outer radii  $R_\mathrm{in}$ and $R_\mathrm{out}$, made of photo-elastic material, and introduce an edge dislocation at its center via a Volterra cut-and-weld procedure (\figref{fig:Figure3}A, see the supplementary materials for more details). When free in 3D, the thin sheet adopts a curved configuration to avoid stresses (see \figref{fig:Figure3}B), similar to defective crystalline membranes and responsive gels that contain an edge dislocation~\cite{seung1988defects,moshe2015geometry}. 

Flattening the sheet between two flat glass plates (\figref{fig:Figure3}C) leads to the buildup of stresses that are fully described by 2D elasticity. 
The stress function in this state is
\begin{equation}
	\psi^\mathrm{dis}_0 (r,\theta) = \frac{Y\,\mathbf{P}\cdot \mathbf{r}}{8\pi }\left[\frac{r^2}{R_\mathrm{in}^2+R_\mathrm{out}^2}  - \frac{R_\mathrm{in}^2R_\mathrm{out}^2}{R_\mathrm{in}^2+R_\mathrm{out}^2} \frac{1}{r^2} -\ln r^2\right].
	\label{eq:dipoleSF}
\end{equation}
Using the stresses derived from this stress function, we calculate the expected photo-elastic pattern  (see Photoelasticity in the supplementary materials), which matches the observed pattern (\figref{fig:Figure3}D). This confirms the validity of the linear elastic solution before the insertion of a crack. 

\begin{figure*}
\centering
\includegraphics[width=0.9\linewidth]{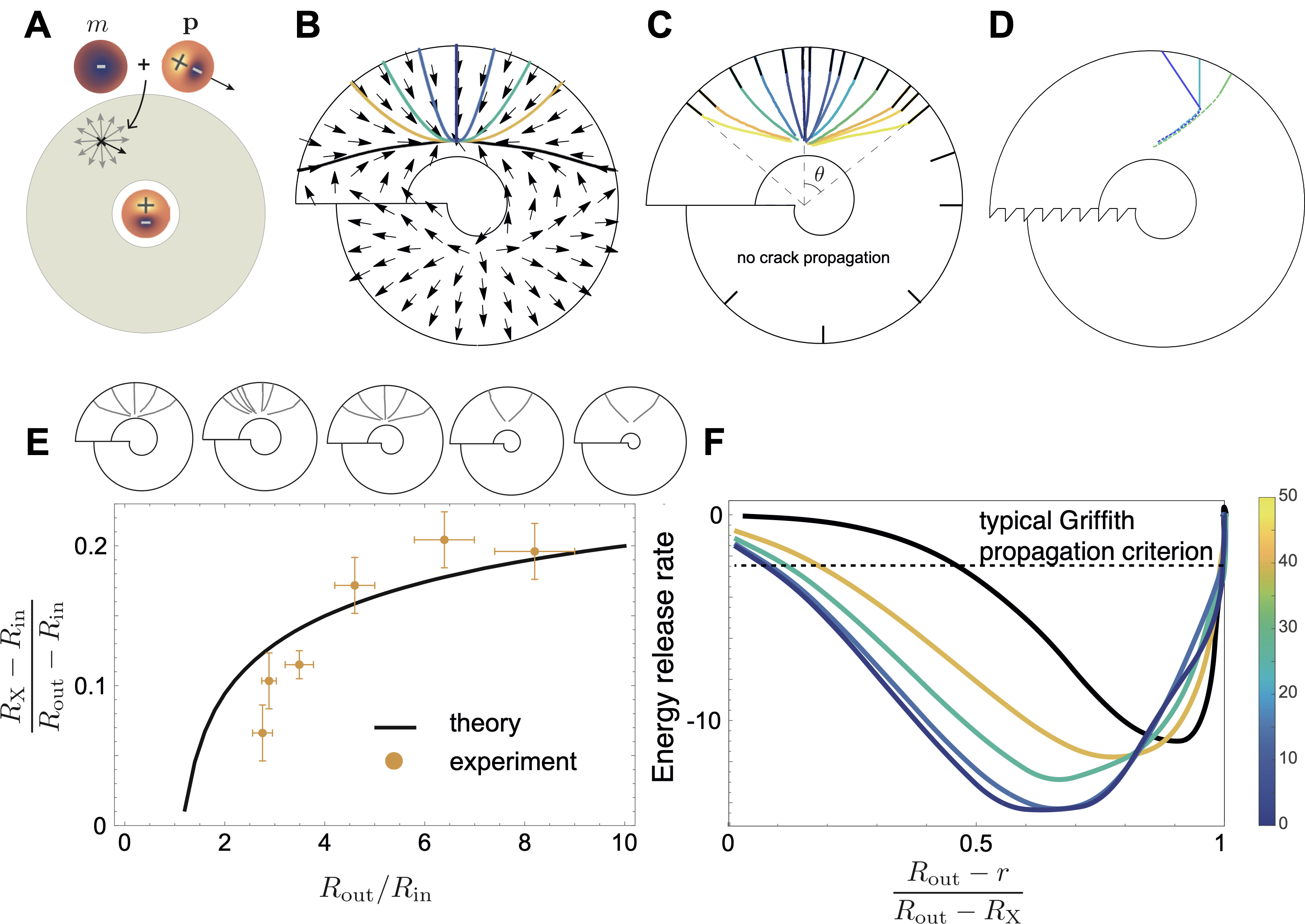}
\caption{\textbf{Cracks trajectory.} \textbf{A}. Schematics of the theoretical approach, where the energy minimizing multipoles $m$ and $\mathbf{p}$ are computed at every point in the dislocated stressed solid. \textbf{B}. Constructing the crack trajectories as streamlines of the dipole field $\mathbf{p}$, we predict that cracks only propagate from one side of the dislocated annulus, and all converge to a single attractor point. \textbf{C}. Experiments also exhibit this remarkable feature. \textbf{D.} Experimental validation of the history-independence prediction. Solid lines  represent different initial crack histories and dashed lines the propagating crack.\textbf{E.} The radial position $R_\mathrm{X}$ of the attractor varies with the geometry of the annulus $R_\mathrm{out}/R_\mathrm{in}$. Our theory quantitatively predicts its position as long as the annulus is thick enough ($R_\mathrm{out}/R_\mathrm{in}>4$). \textbf{F.} Energy release rate in arbitrary units for the trajectories plotted in \textbf{B}, as a function of the distance from the center. The dashed line is a typical Griffith energy release rate; cracks corresponding to curves below the dashed line will propagate while those above will not. Hence, very long initial slits are necessary to have cracks coming from large angles propagate. }
\label{fig:Figure4}
\end{figure*}

To create a crack in the experiment, we cut a notch at the boundary, flattened the dislocated annulus, and let the crack slowly propagate (complete propagation typically takes one hour, consistent with the quasi-static approximation).
\figref{fig:Figure3}E shows the quasi-static temporal evolution of the propagating crack and \figref{fig:Figure4}C summarizes crack trajectories initiated from different locations. The experiment uncovers three characteristic properties of a crack moving near a dislocation. (i) A crack would propagate inwards from the outer boundary only from the side to the right of the Burgers vector and not from the left. (ii) Cracks initiated at different locations propagate in curved trajectories towards the same focal point, located at $\theta = 0$, near, but not at $R_\mathrm{in}$. This convergence to an attractor point has not been observed before, and we are unaware of predictions for such or similar behavior. (iii) Crack trajectories are independent of history, as shown in \figref{fig:Figure4}D, where three different initial cracks ending at the same point lead to the same propagating crack. We now apply the procedure above to obtain the curved trajectories and the attractor point.

To calculate crack trajectories, we note that the result based on Eq.\ref{eq:InfPotential} holds for infinite domains, and requires boundary terms corrections in finite systems. Thus streamlines of $\nabla \psi_0^\mathrm{dis}$ from Eq.\ref{eq:dipoleSF} form an uncontrolled approximation for the crack trajectories. To overcome this we recede back to the full expression of the interaction energy Eq.\ref{eq:int} while still describing the crack interactions using the singular charge approximation. While the interaction energy is obtained analytically at once in terms of $\psi_0^\mathrm{dis}$, the finite domain calls for numerical evaluation of integrals (\figref{fig:Figure4}A, see "Crack Direction and Energy Calculation" in the supplementary materials for a detailed derivation). 
Upon minimization, we find $m$ and $\mathbf{p}$ at every point $\mathbf{x}_c$, as well as the value of the minimal interaction energy $U$.

In \figref{fig:Figure4}B we present the local dipole orientation $\mathbf{p}$ that minimizes $U$, whose streamlines resembles the result in an infinite domain \figref{fig:Figure2}, and accurately recover all experimental observations, from the shape of crack trajectories, through their convergence to a focal point.
Note that these streamlines are \emph{not} identical to the lines of principal stress in the (uncracked) disc (see "Principle Stresses" in the supplementary materials).
To perform a quantitative comparison between theory and experiments, we measure the location of the focal point for various system sizes with different $R_\mathrm{in}$ and $R_\mathrm{out}$.
Theoretically, the focal point is defined as the point at which the energy-minimizing dipole $\mathbf{p}$ becomes insensitive to its orientation, indicating that the derivative of the energy with respect to the dipole orientation vanishes. In \figref{fig:Figure4}E, we present a plot displaying the measured focal points (orange points) alongside the theoretically predicted focal points (solid line) as functions of the system size. Remarkably, we observe excellent agreement between the theory and the experimental results.

Another experimental observation is that cracks propagate from outside only when the initial notch is sufficiently long, as indicated by the black segments in \figref{fig:Figure4}C, and they do not reach the focal point and are stopped slightly earlier. This behavior is explained by considering the energy released during propagation, as derived from the calculated interaction energy.
We plot the energy release rate, that is the derivative of $U$ along the crack path, as a function of crack tip position $\mathbf{x}_c$ for various initial angles \figref{fig:Figure4}F. We also illustrate the fracture energy characterizing the material with a horizontal dashed line. 
This fracture energy line intersect with any curve of potential energy at two points: one close to the outer boundary and another near the focal point. The energy release rate is lower (in absolute value) than the fracture energy beyond these intersection points, confirming the experimental observation described above—cracks are initiated near the outer boundary only beyond a critical initial length and stop near the focal point.
Moreover, the black curves in \figref{fig:Figure4}B do not appear in the experiment, as  \figref{fig:Figure4}C shows. We explain this by the fact that the corresponding energy release rate seen, expressed by the black curve in \figref{fig:Figure4}F, is lower (in absolute value) than the fracture energy. Therefore the crack does not propagate. 
Even though the presented model describes quasi-static cracks, its treatment of the energy release rate suggests a possible description of time-dependent propagation. As shown in Fig. 4F, the energy release rate initially increases in absolute value before reaching a peak and subsequently decreasing. This trend indicates that the force acting on the crack first strengthens, leading to acceleration, and then weakens, causing deceleration until the energy reaches the Griffith threshold, at which point the crack stops. Importantly, the system remains quasi-static, as the experiment shown in the supplementary video took approximately an hour to complete.
This agreement further supports the validity of LEFM and of our theory.

\section*{Discussion}

In summary, our work presents a novel framework for characterizing propagating cracks, which aligns seamlessly with experimental observations. Its primary advantage lies in simplifying the description of cracks to point-like objects governed by principles akin to Newtonian particle mechanics. Our findings demonstrate that crack propagation is primarily governed by the forces acting on its monopole and dipole components, in addition to the torque acting on the dipole. Notably, our framework predicts that crack trajectories are independent of the crack’s history, depending only on its current tip position and the background stress field. This is in contrast to conventional views in fracture mechanics, where the full crack path is often thought to be highly dependent on past propagation.

The predicted orientation of the dipole is established through energy minimization, which is equivalent to maximizing the energy release rate. In the propagation phase, adhering to the principle of local symmetry, the crack advances in the direction of maximum tensile stress, aligning with the local orientation of the dipole.
We have employed Griffith's criterion to ascertain whether a crack propagates, integrating it into our methodology through an energy minimization procedure. An intriguing insight within our framework is the possible interpretation of this criterion as an effective static friction force. This observation paves the way for additional analogies with particle mechanics to explore crack propagation, including formulating an equation of motion for the crack tip's position and orientation within over-damped dynamics. 

A key advantage of our framework, beyond its analytical power, is its ability to predict crack trajectories directly from the pre-stressed state, eliminating the need for the iterative approaches traditionally considered inherent to LEFM. This feature becomes particularly valuable in complex setups, such as systems with displacement-controlled boundary conditions or low symmetry, where conventional methods face significant challenges. In such cases, our framework can be efficiently implemented numerically, streamlining the prediction of crack paths even in intricate geometries.

While our work focuses on 2D planar elasticity, where the crack tip is a point-like object and the elastic charge is a scalar quantity that describes its interaction with the background, generalizing our approach to 3D fracture mechanics presents new challenges. In three dimensions, the crack tip is no longer a point but a string-like object, and elastic charges naturally generalize to line charges, analogous to disclination and dislocation lines in elasticity theory. The primary complexity arises in describing the dynamics of an evolving string rather than a point-like object. Extending our theoretical framework to 3D and investigating the evolution of these string-like structures is an ongoing area of research.

A central assumption in our work is that the background stress, represented by $\psi_0$, remains constant during crack propagation, enabling the minimization procedure for the crack’s monopole and dipole. This approximation holds when driving stresses originate from static remote sources. However, it may break down when a crack interacts with another crack, especially when one crack is close to the tail of the other.
To account for such interactions, our framework can be refined by replacing the singular charge approximation with a fixed charge density approximation. In this model, the crack maintains a fixed charge distribution regardless of its shape, consisting of a negative singular charge at the tip and a positively charged tail that decays following a characteristic power law. While interactions with remote stress sources can still be approximated using singular charges, crack-crack interactions require considering the full charge distribution along the tail. This refinement provides a more accurate description of interacting cracks and lays the groundwork for future research.

Lastly, it's important to note that our current discussion is confined to analysis of crack propagation in a Hookean elastic material. However, our charges formulation can be applied to any elastic model and propagation criteria. Therefore, we have plans to extend our approach more comprehensive elasticity models, such as Neo-Hookean \cite{li2023crack} and screened solids \cite{livne2023geometric}. The latter is of particular interest for studying crack trajectories in solids with relaxational degrees of freedom, from glasses, to fabrics, and more.

\bibliographystyle{Science}

\section*{Acknowledgments}
 This work was supported by Israel Science Foundation Grant No. 1441/19, by NSF-BSF grant No. 2020739, and by the International Research Project ``Non-Equilibrium Physics of Complex Systems'' (IRP-PhyComSys, France-Israel).
\section*{Supplementary materials}
Materials and Methods\\
Supplementary Text\\
Figs. S1 to S3\\
Tables S1 to S4\\

\end{document}



\baselineskip24pt


\maketitle 

\tableofcontents

\section{Materials and Methods}

\subsection*{Sample preparation}
Samples are made as follows. First, flat sheets of constant thickness are made of either a photoelastic polyurethane (Clearflex 40 from Smooth On) or a silicone (Rubber Glass from Smooth On) rubber. Equal quantities of catalyst and base liquids are thoroughly mixed and then poured onto a flat glass plate sprayed with a release agent (Ease-Release from Mann). A covering glass plate with spacers of 1mm is then placed on top of the liquid to ensure a constant thickness of the rubber sheet. After curing at room temperature, the 1mm-thick sheets are laser cut with a pattern made of two concentric Archimedes spirals connected by a jigsaw line (see Fig.~2A in the main article). A shear is imposed along the jigsaw line to close the structure, yielding a dislocation defect at its center. A small amount of the same mixed liquid is used as a glue. After curing, a notch is initiated perpendicular to the boundary of the sheet. The structure is then covered with talc powder to reduce friction and flattened between two glass plates.

\subsection*{Experimental setup}
The structure, flattened between two glass plates, is placed above a source of polarized light (a white screen facing up). A top-view camera equipped with a cross-polarizer records the crack propagation. With this setup, stresses may be visualized thanks to the photoelastic property of the polyurethan rubber (see next section for the details). It typically takes one hour in the case of polyurethane rubber and a few seconds for silicone rubber for the crack to fully propagate. Once the propagation has stopped, the structure is cut along the jigsaw line to release the geometrical constraint. The crack in the undeformed configuration is highlighted with a permanent marker. A top-view picture is then analyzed to get the crack trajectory (Fig.~3A in the main article).

\subsection*{Photoelasticity}
\label{app:photoelasticity}
Polyurethane rubber (Clearflex 40 from Smooth On) is known to exhibit strong photoelastic properties ~\cite{daniels2017photoelastic}. This means that the material, when submitted to mechanical stress, become birefringent: the material has two distinctive refraction indexes along the principal stresses direction, the magnitude of the refractive indices at each point in the material being directly related to the state of stress at that point. 
Hence, when a linearly polarized light -- i.e., waves that all have the same orientation as their electric fields -- passes through the medium, it rotates its polarization. Indeed, its electromagnetic wave components along the two principal stress directions experience a different refractive index due to the birefringence. The difference in the refractive indices leads to a relative phase retardation $\Delta$ between the two components, that is proportional to the principal stresses difference and reads:

\begin{equation}
    \Delta=C\frac{2\pi t}{\lambda}\left(\sigma_1-\sigma_2\right),
\end{equation}
where $t$ is the sheet thickness, $\lambda$ the light wavelength, ($\sigma_1,\sigma_2$) the principal stresses and $C$ the stress-optic coefficient, that slightly depends on the wavelength.
The light then passes through the final orthogonal polarizer fixed on the camera lens. The final intensity of light thus depends on the phase retardation $\Delta$ and reads:
\begin{equation}
    I(x,y)=I_0\sin^2\left(\frac{\Delta}{2}\right)
    \label{eq:intensity}
\end{equation}

Note that if the principal stress of the sample happens to be aligned with the polarizer, it will not be affected by the photoelastic response of the material. Therefore, the fringe pattern also depends on the relative angles between the polarizer (fixed) and the principal stress. This explains the dark lines in Fig.~2D in the main text (and in Fig.~S1C). In the case of white light, each wavelength thus rotates by a different amount, leading to the iridescent patterns shown in Fig.~2 in the main text and in Fig.~S1C.

In order to calibrate our experiments, we stretch a beam of the same thickness $t=1$ mm made of the same polyurethane rubber as in the experiment and illuminated with a monochromatic red polarized backlight. Upon stretching, the beam appearance alternates between dark and light as the principal stress difference $\sigma_1-\sigma_2=Y/(1+\nu)[\epsilon_1-\epsilon_2]$ is increased (Fig.~S1A). From this observation, we may deduce the  stress-optic coefficient $C$ by the successive fringes (Fig.~S1B). In order to quantitatively validate our approach in a simple configuration, we flatten a cone (i.e., a disclination, Fig.~S1C), yielding an axisymmetric pattern. Analyzing the intensity of each color channel (red, green and blue) along a radial line (Fig.~S1D) yields the evolution of the principal stress difference up to a constant. In order to fix this constant, we find the location where all the 3 color channels exhibit an in-phase local minimum of intensity: it corresponds to the place where the material is not birefringent, and hence where $\Delta\sigma=0$ (Fig.~S1D). Shifting accordingly the experimental data, we get a very good agreement with the classical in-plane theory for a disclination defect~\cite{seung1988defects} (Fig.~S1E).
\begin{figure*}
\centering
\includegraphics[width=0.8\linewidth]{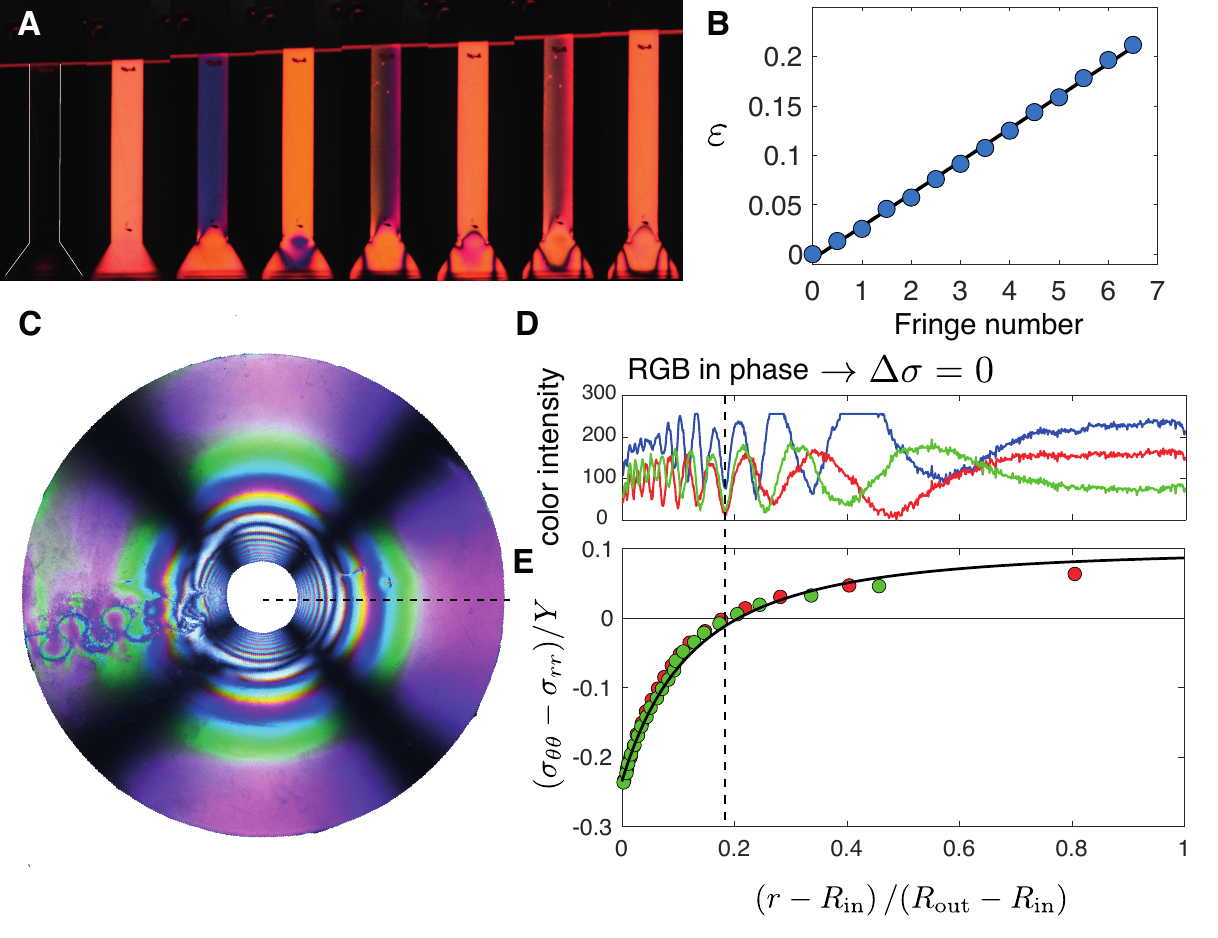}
\caption*{Fig. S1: \textbf{Photoelasticity calibration.} \textbf{A}. Alternation of dark and light appearance of a beam made of polyurethane elastomer and lightened by a monochromatic red polarized source as it is gradually stretched. \textbf{B.} Imposed strain as a function of the number of fringe observed. \textbf{C.} Axisymmetric photoelastic pattern of a flattened cone -- i.e., a disclination --, with $R_\mathrm{in}=18$ mm, $R_\mathrm{out}=90$ mm and an angle deficit of 60 $^\circ$. \textbf{D.} Color intensity as a function of the radial distance for each color channel (red, green and blue). \textbf{E.} Dimensionless principal stress difference as a function of the radial distance. The circles correspond to the experimental points extracted form the color intensity profile for the red and green channels respectively. The dark line corresponds to the classical in-plane theory~\cite{seung1988defects}.  }
\label{fig:fig2}
\end{figure*}

We now consider the case of a dislocation, presented in Fig.~2D. In order to reproduce the light pattern from the theory, we simply color the sheet as a function of the principal stress difference (see the Section 6 for a detailed derivation), using  Eq.~\ref{eq:intensity} for the red, green and blue channels separately.

\subsection*{Crack Direction and Energy Calculation}
In the experiments presented in the paper, the system is relatively small compared to the crack itself, and therefore the boundary effects are important. Consequently, we perform a numerical integration of the energy density.

To find the optimal orientation for every point $\left(x,y\right)$, we only calculate the interaction of the dipole and minimize it with respect to its orientation, i.e.
\begin{equation}
    \min\int A\sigma_\textit{0}\sigma_\textit{d}\left(\mathbf{r_0},\mathbf{p}\right)dS\;,
\end{equation}
where $\sigma_\textit{0}$ is the background stress and $\sigma_\textit{d}\left(\mathbf{r_0},\mathbf{p}\right)$ is the stress of a point dipole at point $\mathbf{r_0}$ with a dipole vector $\mathbf{p}$, having the size $p$ and angle $\theta$. Other terms were neglected because they do not affect the orientation.

To calculate the energy release rate, the interaction between the monopole and the background and the self-interaction, i.e., the energy required to create a monopole and a dipole, were calculated, i.e.,
\begin{equation}
    \min\int A_{\alpha\beta\gamma\delta}\sigma_\textit{0}^{\alpha\beta}\left[\sigma_\textit{d}^{\gamma\delta}\left(\mathbf{r_0},\mathbf{p}\right)+\sigma_\textit{m}^{\gamma\delta}\left(\mathbf{r_0},q\right)\right]+A_{\alpha\beta\gamma\delta}\left[\sigma_\textit{d}^{\alpha\beta}\left(\mathbf{r_0},\mathbf{p}\right)\sigma_\textit{d}^{\gamma\delta}\left(\mathbf{r_0},\mathbf{p}\right)+\sigma_\textit{m}^{\alpha\beta}\left(\mathbf{r_0},q\right)\sigma_\textit{m}^{\gamma\delta}\left(\mathbf{r_0},q\right)\right]dS\;,
\end{equation}
where $\sigma_\textit{m}\left(\mathbf{r_0},q\right)$ is the stress of a point monopole at point $\mathbf{r_0}$
The self-interaction of the background was neglected because it contributes a constant and, therefore, does not affect the energy release rate.

The stress functions from which the stresses above are derived are as follows:
\begin{subequations}
    \begin{align}
        \psi_0\left(r,\theta\right)& = \frac{Y\,\mathbf{P}\cdot \mathbf{r}}{8\pi }\left(\frac{r^2}{r_\text{in}^2+r_\text{out}^2}  - \frac{r_\text{in}^2r_\text{out}^2}{r_\text{in}^2+r_\text{out}^2} \frac{1}{r^2} -\ln r^2\right)\;,\\
        \psi_d\left(r,\theta\right)& = \frac{Y\,\mathbf{p}\cdot \left(\mathbf{r}-\mathbf{r}_0\right)}{8\pi }\ln {\left(\mathbf{r}-\mathbf{r}_0\right)^2}\;,\\
        \psi_m\left(r,\theta\right)& =\frac{Yq\left(\mathbf{r}-\mathbf{r}_0\right)^2}{8\pi}\ln{\left(\mathbf{r}-\mathbf{r}_0\right)^2}\;.
    \end{align}
\end{subequations}


\section{Equivalence of charge and boundary conditions}
\label{App:equivalence}
\subsubsection*{Electric monopole density}
Consider an electrically conductive material. Its energy can be written as
\begin{equation}
    E=\frac{1}{2}\iint\left[\left(\nabla\varphi\right)^2\right]dS-\oint_{\partial\Omega}\varphi\rho d\ell
\end{equation}
with $\phi$ the electric potential and $\rho$ the charge density. The charge is concentrated on the surface because the material is conductive. We minimize the energy for the potential and the charge density and get the following equations:
\begin{subequations}
    \begin{gather}
        \frac{\delta E}{\delta\varphi\left(x\right)}=-\iint\left[\nabla^2\varphi \left(x\right)\delta\left(x-x'\right)\right]dS+\oint\left[\nabla\varphi\cdot \hat{n}-\rho\right]\delta\left(x-x'\right)d\ell\\
        \frac{\delta E}{\delta\rho}=-\oint_{\partial\Omega}\varphi\delta\left(x-x'\right)d\ell
    \end{gather}
\end{subequations}
where $\frac{\delta E}{\delta\varphi\left(x\right)}$ is the variational derivative of the energy with respect to the electric potential. We find that the electric potential and charge density that minimizes the energy satisfies the Laplace equation with the known boundary condition
\begin{equation}
    \Vec{E}\cdot\hat{n}=\rho\;.
\end{equation}

\subsubsection*{Electric dipole density}
Because the monopole is a conserved quantity, it is important to consider a dipole distribution. In the case of a dipole distribution, the energy takes the form
\begin{equation}
    E=\frac{1}{2}\iint\left[\left(\nabla\varphi\right)^2\right]dS-\oint_{\partial\Omega}\varphi\nabla\cdot\Vec{P} d\ell,\;\Vec{P}=p\Vec{t}
\end{equation}
where the dipoles are parallel to the boundary.
We minimize the energy with respect to the potential and the dipole density and get the following equations:
\begin{subequations}
    \begin{gather}
         \frac{\delta E}{\delta\varphi\left(x\right)}=-\iint\left[\nabla^2\varphi \left(x\right)\delta\left(x-x'\right)\right]dS+\oint\left[\nabla\varphi\cdot \hat{n}-\nabla\cdot\Vec{P}\right]\delta\left(x-x'\right)d\ell\\
         \frac{\delta E}{\delta p}=\oint_{\partial\Omega}\nabla\varphi\cdot\Vec{t}\delta\left(x-x'\right)d\ell
    \end{gather}
\end{subequations}
The electric potential must satisfy the Laplace equation with the boundary conditions
\begin{subequations}
    \begin{gather}
        \Vec{E}\cdot\hat{n}=\nabla\cdot\Vec{P}\\
        \Vec{E}\cdot\Vec{t}=0\;.
    \end{gather}
\end{subequations}

\subsubsection*{Elastic quadrupole density}
The main equation of elasticity, which is analogous to the Laplace equation in electrostatics, is the biharmonic equation, The equivalent of the electric charge, i.e., the source term of the equation is elastic defects or curvature \cite{seung1988defects}
\begin{equation}
    \frac{1}{Y}\Delta\Delta\psi=K_d
\end{equation}
The source term can be expanded in a multipole expansion
\begin{equation}
    K_d=M\left(x\right)+\nabla_\alpha P^\alpha\left(x\right)+\nabla_{\alpha \beta}Q^{\alpha\beta}\left(x\right)+...\;,
\end{equation}
with M, P, and Q distributions of disclinations, dislocations, and quadrupole.
P is defined by
\begin{equation}
    P^\alpha=\varepsilon^{\alpha\beta}b_{\beta}
\end{equation}
with b being the Burger's vector, while Q is defined by
\begin{equation}
Q^{\alpha\beta}=\varepsilon^{\alpha\mu}\varepsilon^{\beta\nu}q_{\mu\nu}
\end{equation}
with q being a local change in the reference state.
Like the dipole in electrostatics, the quadrupole is the lowest non-conserved moment in elasticity. Therefore, we want to find the quadrupole distribution that minimizes the elastic energy. The energy functional takes the form
\begin{equation}
    E=\frac{1}{2}\iint \tilde{A}^{\alpha\beta\gamma\delta}\nabla_{\alpha\beta}\psi\nabla_{\gamma\delta}\psi dS-\oint\psi \nabla_{\alpha\beta}Q^{\alpha\beta}d\ell
\end{equation}
where we define
\begin{equation*}
    \tilde{A}^{\alpha\beta\gamma\delta}=A_{\mu\nu\rho\sigma}\varepsilon^{\alpha\mu}\varepsilon^{\beta\nu}\varepsilon^{\gamma\rho}\varepsilon^{\delta\sigma}\;,
\end{equation*}
not to be confused with $A^{\alpha\beta\gamma\delta}=g^{\alpha\mu}g^{\beta\nu}g^{\gamma\rho}g^{\delta\sigma}A_{\mu\nu\rho\sigma}$.
$A^{\alpha\beta\gamma\delta}$ is the elastic tensor, $\psi$ is the Airy stress function, $\varepsilon^{\alpha\beta}$ is the Levi-Civita symbol and $g^{\alpha\beta}$ is the metric.

From the variation of the stress function, we get
\begin{equation}
    \frac{\delta E}{\delta \psi}=\iint \tilde{A}^{\alpha\beta\gamma\delta}\nabla_{\alpha\beta\gamma\delta}\psi\delta\left(x-x'\right)dS-\oint \left[2\tilde{A}^{\alpha\beta\gamma\delta}\nabla_{\alpha\beta\gamma}\psi n_\delta+\nabla_{\alpha\beta}Q^{\alpha\beta}\right]\delta\left(x-x'\right)d\ell
\end{equation}
From the bulk term, we get
\begin{equation}
    \tilde{A}^{\alpha\beta\gamma\delta}\nabla_{\alpha\beta\gamma\delta}\psi=\frac{1}{Y}\nabla^4\psi=0
\end{equation}
and from the boundary term, we get
\begin{equation}
    \frac{1}{Y}\nabla_n\nabla^2\psi=\frac{1}{Y}\nabla_n P=-\nabla^2Q
\end{equation}
where $P$ is the pressure and $\nabla_n$ is the derivative in the normal direction.

From the derivative with respect to the quadrupole, we get
\begin{equation}
        \frac{\delta E}{\delta q}=-\oint\sigma^{\mu\nu}\delta q_{\mu\nu}
\end{equation}

The most general form of $q_{\alpha\beta}$ is $q_{\alpha\beta}=u_\alpha v_\beta+u_\beta v_\alpha$, and to get the appropriate boundary condition set $u_\mu=n_{\mu}$ with $n_{\mu}$ being the normal vector.

\begin{equation}
    \sigma^{\mu\nu}\delta q_{\mu\nu}=\sigma^{\mu\nu}\left(n_\mu\delta v_\nu+n_{\nu}\delta v_\mu\right)=2\sigma^{\mu\nu}n_{\mu}\delta v_\nu
\end{equation}
which is the known boundary condition
\begin{equation}
    \sigma\cdot\hat{n}=0
\end{equation}

The vector $v$ can be generally written as $\frac{1}{2}\left(\alpha n+\beta t\right)$, and therefore, the general quadrupole moment that satisfies the boundary conditions is
\begin{equation}
    q_{\mu\nu}=n_\mu v_\nu+n_{\nu} v_\mu=\alpha n_\mu n_\nu+\beta\frac{1}{2}\left(n_\mu t_\nu+n_\nu t_\mu\right)
\end{equation}

\section{Charge Density}
\label{App:charge_density}
Consider a mode I crack defined by the geometry $\{(x,y):-l_c/2<x<l_c/2 \;\land\; y=0\}$, subjected to constant uniaxial load in the y direction $\sigma_\infty$. Upon measuring distances in units of $l_c$ and stresses in units of $Y$, the stress function of this problem is
\begin{equation}
    \psi\left(z,\Bar{z}\right)=\frac{1}{2}\left(\Bar{z}\phi\left(z\right)+\eta\left(z\right)+z\overline{\phi\left(z\right)}+\overline{\eta\left(z\right)}\right)
\end{equation}
with
\begin{subequations}
    \begin{align}
        \phi\left(z\right)&=\frac{1}{4} \sigma_\infty  \left(2 \sqrt{z^2-1}-z\right)\\
        \eta'\left(z\right)&=\chi\left(z\right)=\frac{1}{2} \sigma_\infty  \left(z-\frac{1}{\sqrt{z^2-1}}\right)
    \end{align}
\end{subequations}
The stress function $\psi$ satisfies the equation
\begin{equation}
    \frac{1}{Y}\Delta\Delta\psi\left(z,\Bar{z}\right)=0\;.
\end{equation}
with traction-free boundary conditions along the crack. 
We showed in Equivalence of charge and boundary conditions in the Supplementary Material that finding the charge density $\rho$ that satisfies the bi-harmonic equation is equivalent to satisfying boundary conditions. Therefore, we find the equivalent charge density for the crack problem.

\begin{equation}
    \frac{1}{Y}\Delta\Delta\psi\left(z,\Bar{z}\right)=K\;.
\end{equation}

To calculate the charge density along the crack, we use the same method used to calculate the surface charge density in electrostatics, i.e., calculate the jump in the electric field between the two sides of the medium. The equivalent of the electric potential, the scalar function that satisfies the Laplace equation, is the Laplacian of the Airy stress function:
\begin{equation}
    \Delta\psi\left(z,\Bar{z}\right)=\frac{1}{2} \sigma_\infty 
   \left(\frac{z}{\sqrt{\left(z-1\right)\left(z+1\right)}}+\frac{\Bar{z}}{\sqrt{\left(\Bar{z}-1\right)\left(\Bar{z}+1\right)}}-1\right)
\end{equation}
The derivative in the direction normal to the crack faces gives
\begin{equation}
    \nabla_{n}\Delta\psi\left(z,\Bar{z}\right)=\sigma_\infty\frac{i \left(\sqrt{z^2-1} z^2-\sqrt{z^2-1}-\Bar{z}^2 \sqrt{\Bar{z}^2-1}+\sqrt{\Bar{z}^2-1}\right)}{2 \left(z^2-1\right)^{3/2}
   \left(\Bar{z}^2-1\right)^{3/2}}
\end{equation}
We take $z=x+i\epsilon$ with $\epsilon\rightarrow 0^+$. Outside the crack, this expression vanishes. However, along the crack, $\sqrt{\bar{z}^2-1}=\sqrt{z^2-1}$, which gives the dimensionless charge density along the crack
\begin{equation}
    \tilde{\rho}\left(x\right)=\begin{cases}
        \frac{\sigma_\infty}{\left(1-x^2\right)^{3/2}},&-1<x<1\\
        0,& \text{otherwise}
    \end{cases}
\end{equation}
Bringing back the dimension $l_c$ and $Y$ we find
\begin{equation}
	\rho\left(x\right)=\begin{cases}
		\frac{l_c^2 \, \sigma_\infty}{Y \left(l_c^2-x^2\right)^{3/2}},&-1<x<1\\
		0,& \text{otherwise}
	\end{cases}
\end{equation}
This charge distribution is equivalent to the dipole (Burger's vector) distribution given in \cite{landau1959course}, with the charge distribution being the divergence of the dipole distribution, according to the multipole expansion.

\subsubsection*{Center of charge}
Let us look at the right half of the crack, i.e., $0<x<1$. The center of charge is then calculated as
\begin{equation}
    \lim _{\epsilon\rightarrow 1^+}\frac{\int_0^\epsilon \rho\left(x\right)x dx}{\int_0^\epsilon \rho\left(x\right)dx}=\lim _{\epsilon\rightarrow 1^+} \frac{1-\sqrt{1-\epsilon ^2}}{\epsilon }=1\;.
\end{equation}
Therefore, the charge center where we perform the multipole expansion is at the crack's tips.

\section{Asymptotic Behaviour}
\label{App:asymptotic}
In the classical formulation of linear elasticity, the Airy stress function satisfies the Biharmonic equation
\begin{equation}
    \Delta\Delta\psi=0\;.
\end{equation}
We solve the equation using the separation of variables. Guessing a solution of the form $\psi\left(r,\theta\right)=r^{\lambda+1}\Theta\left(\theta\right)$ \cite{williams1957stress}, we get the following equation for the angular part
\begin{equation}
    \Theta ^{(4)}(\theta )+2 \left(\lambda ^2+1\right) \Theta ''(\theta )+\left(\lambda
   ^2-1\right)^2 \Theta (\theta )=0
\end{equation}
and get the general solution
\begin{equation}
    \psi\left(r,\theta\right)=r^{\lambda+1}\left[A_1 \cos{\left(\lambda+1\right)\theta}+A_2 \cos{\left(\lambda-1\right)\theta}+B_1 \sin{\left(\lambda+1\right)\theta}+B_2 \sin{\left(\lambda-1\right)\theta}\right]
\end{equation}
Our problem is a straight mode I crack where the origin is at the crack tip and $\theta = 0$ is the direction in which the crack points. Therefore, our solution should satisfy $\psi\left(r,\theta\right)=\psi\left(r,-\theta\right)$ and therefore we are left with
\begin{equation}
    \psi\left(r,\theta\right)=r^{\lambda+1}\left[A_1 \cos{\left(\lambda+1\right)\theta}+A_2 \cos{\left(\lambda-1\right)\theta}\right]
\end{equation}
The traction-free boundary conditions are
\begin{equation}
    \sigma^{\theta\theta}\left(r,\pm\pi\right)=\sigma^{r\theta}\left(r,\pm\pi\right)=0
\end{equation}
\begin{subequations}
    \begin{gather}
        \sigma^{\theta\theta}=\frac{\partial^2 \psi\left(r,\theta\right)}{\partial r^2}=\lambda  \left(\lambda +1\right) r^{\lambda -1} \left[A_1 \cos \left(\lambda +1\right)\theta+A_2
   \cos \left(\lambda -1\right)\theta\right]\\
        \sigma^{r\theta}=-\frac{\partial }{\partial r}\left(\frac{1}{r}\frac{\partial \psi\left(r,\theta\right)}{\partial \theta}\right)=\lambda  r^{\lambda -1} \left[A_1 \left(\lambda +1\right) \sin\left(\lambda +1\right)\theta+A_2
   \left(\lambda -1\right) \sin \left(\lambda -1\right)\theta\right]
    \end{gather}
\end{subequations}

After the substation of $\theta=\pm\pi$, we get the following equations.

\begin{subequations}
    \begin{gather}
        \left(A_1+A_2\right) \cos \left(\lambda +1\right)\pi=0\\
     \left[  A_1 \left(\lambda +1\right) +A_2
   \left(\lambda -1\right)\right] \sin \left(\lambda +1\right)\pi=0
    \end{gather}
\end{subequations}

The set of equations has two possible solution:
\begin{subequations}
    \begin{gather}
        \lambda=n,\;A_2=-A_1\\
        \lambda=\frac{1+2n}{2},\;A_2=\frac{1+\lambda}{1-\lambda}A_1
    \end{gather}
\end{subequations}
The displacement continuity requires the power of $r$ in the stress function to be larger than one \cite{williams1952stress}.

We are looking for the asymptotic behavior and the lowest power of $r$, i.e., $\lambda=\frac{1}{2}$. Therefore, the asymptotic stress function takes the following form.

\begin{equation}
    \psi_{\text{Asym}}\left(r,\right)=A r^{3/2}\left[\cos{\left(\frac{3}{2}\theta \right)}+3\cos{\left(\frac{1}{2}\theta\right)}\right]\;,
\end{equation}
where $A$ is determined by the applied traction far from the crack.

To find the asymptotic charge distribution, we calculate the jump in the gradient of the Laplacian, the way performed in calculating surface charge density in electrostatics. The derivative of the Laplacian normal to the crack, i.e., the $\theta$ direction, is 
\begin{equation}
    \frac{\partial}{\partial_\theta}\Delta\psi_{\text{Asym}}\left(r\right) =-A\frac{3 \sin \left(\frac{\theta }{2}\right)}{r^{3/2}}
\end{equation}
Indeed, the derivative is discontinuous along the crack, where $\theta=\pm\pi$, which gives the asymptotic charge density
\begin{equation}
    \rho_{\text{Asym}}\propto r^{-3/2}
\end{equation}
\section{Interactions in finite domains}
\label{App:finite_domain}
The energy functional for Hookean solids is
\begin{equation}
    E=\frac{1}{2}\int A_{\alpha\beta\gamma\delta}\sigma^{\alpha\beta}\sigma^{\gamma\delta}\sqrt{\bar{g}}dS
\end{equation}
where $\bar{g}$ is the determinant of the reference metric. One can write the stress as derivatives of the scalar Airy stress potential $\psi$
\begin{equation}
    \sigma^{\alpha\beta}=\frac{1}{\sqrt{g}\sqrt{\bar{g}}}\varepsilon^{\alpha\mu}\varepsilon^{\beta\nu}\nabla_\mu\bar{\nabla}_\nu\psi,\;
\end{equation}
which, under the assumptions of linear elasticity, become
\begin{equation}
    \sigma^{\alpha\beta}=\frac{1}{\bar{g}}\varepsilon^{\alpha\mu}\varepsilon^{\beta\nu}\bar{\nabla}_{\mu\nu}\psi\;.
\end{equation}
Therefore, one can write the energy functional as
\begin{equation}
   E=\frac{1}{2}\int \frac{1}{\bar{g}^2}A_{\alpha\beta\gamma\delta}\varepsilon^{\alpha\mu}\varepsilon^{\beta\nu}\varepsilon^{\gamma\rho}\varepsilon^{\delta\sigma}\left(\bar{\nabla}_{\mu\nu}\psi \right)\left(\bar{\nabla}_{\rho\sigma}\psi\right)\sqrt{\bar{g}}dS
\end{equation}
One may integrate this equation twice by part to get
\begin{equation} 
   E=
   \frac{1}{2}\int \frac{1}{\bar{g}^2}A_{\alpha\beta\gamma\delta}\varepsilon^{\alpha\mu}\varepsilon^{\beta\nu}\varepsilon^{\gamma\rho}\varepsilon^{\delta\sigma}\left(\bar{\nabla}_{\mu\nu\sigma\rho}\psi \right)\psi\sqrt{\bar{g}}dS+\mathcal{B}
\end{equation}

where $\mathcal{B}$ is the boundary term
\begin{equation}
    \mathcal{B}=\frac{1}{2}\oint \frac{1}{\bar{g}^2}A_{\alpha\beta\gamma\delta}\varepsilon^{\alpha\mu}\varepsilon^{\beta\nu}\varepsilon^{\gamma\rho}\varepsilon^{\delta\sigma}\left[\left(\bar{\nabla}_{\mu\nu}\psi \right)\left(\bar{\nabla}_{\rho}\psi\right)-\left(\bar{\nabla}_{\mu\nu\rho}\psi \right)\psi\right]\sqrt{\bar{g}}d\ell_{\sigma}  
\end{equation}
By rewriting the bulk term \cite{moshe2015elastic}, the energy functional takes the form:
\begin{equation} 
   E=
   \frac{1}{2}\int \psi\bar{K}_G dS+\mathcal{B}
\end{equation}
with $\bar{K}_G$ being the Gaussian curvature of the reference metric.

\section{Principal stresses do not yield accurate crack trajectories}
\label{App:principal}
We want to compare the principal stresses of the system to the optimal direction of a dipole in the system, i.e., the potential path of the crack. 
Fig.1 compares the two cases. The dipole orientation, presented in the left panel, is calculated by minimizing the elastic energy with respect to its orientation. In the right panel, we show the direction of the largest principal stress, i.e., the eigenvector with the largest eigenvalue of the stress tensor.
One can see that the analysis of the principal stresses does not reveal the crack paths. This is most evident while looking at the focal point in the left panel but is absent in the principal stress plot.
\\
\\
To calculate the principal stresses, we must first solve the Biharmonic equation for the case of a dipole charge, that is

\begin{equation}
    \frac{1}{Y}\Delta\Delta\psi=\mathbf{p}\cdot\nabla\delta\left(\mathbf{r}\right)
\end{equation}
with the solution for the stress function being
\begin{equation}
    \frac{Y}{8\pi}\mathbf{p}\cdot\mathbf{r}\left(A r^2+B r^{-2}-\ln{r^2}\right)
\end{equation}
where we choose A and B to satisfy the boundary conditions \cite{seung1988defects, moshe2015elastic}.

One can derive the stress tensor from the stress function by
\begin{equation}
    \sigma^{\mu\nu}=\varepsilon^{\mu\alpha}\varepsilon^{\nu\beta}\nabla_{\alpha\beta}\psi
\end{equation}
with $\varepsilon^{\mu\alpha}$ being the Levi-Civita symbol \cite{landau1959course,moshe2014plane}.

To get the principal stresses from the stress tensor, one needs to diagonalize it. This gives the following principal stresses (eigenvalues)

\begin{subequations}
    \begin{align}
         \sigma_1 &= \frac{1}{2} \left(\sigma^{xx} + \sigma^{yy}\right)  + \sqrt{(\sigma^{xx} - \sigma^{yy})^2 + 4 \left(\sigma^{xy}\right)^2}\\
    \sigma_2 &= \frac{1}{2} \left(\sigma^{xx} + \sigma^{yy}\right)  - \sqrt{(\sigma^{xx} - \sigma^{yy})^2 + 4 \left(\sigma^{xy}\right)^2}
    \end{align}
\end{subequations}

with the corresponding eigenvectors

\begin{equation}
    \mathbf{v_1}=\left(
\begin{array}{c}
 \frac{\sqrt{(\sigma^{xx}-\sigma^{yy})^2+4 \left(\sigma^{xy}\right)^2}+\sigma^{xx}-\sigma^{yy}}{2 \sigma^{xy}} \\
 1 \\
\end{array}
\right)\;,
\quad
\mathbf{v_2}=\left(
\begin{array}{c}
 -\frac{\sqrt{(\sigma^{xx}-\sigma^{yy})^2+4 \left(\sigma^{xy}\right)^2}-\sigma^{xx}+\sigma^{yy}}{2 \sigma^{xy}} \\
 1 \\
\end{array}
\right)\;.
\end{equation}

\clearpage

\section*{Fig. S1.}
\begin{figure*}[h]
\centering
\includegraphics[]{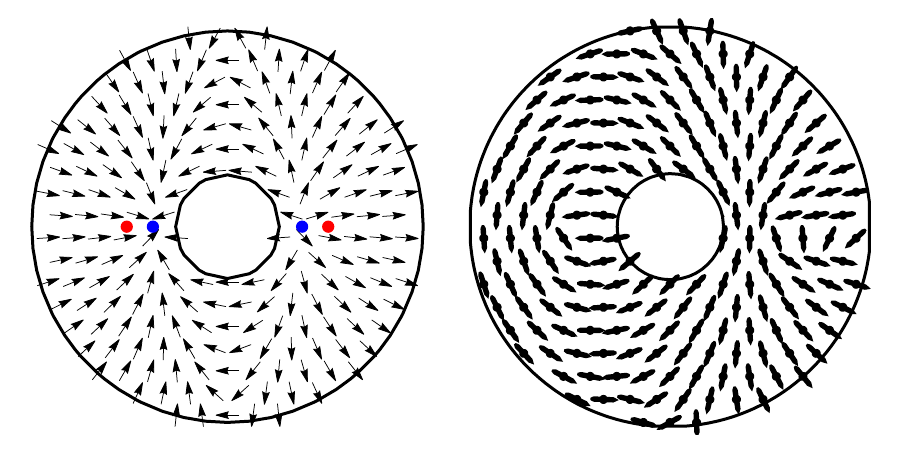}
\caption{\textbf{Principal stresses and focal point} A comparison between the principal background stresses (right) and the preferred dipole orientation (left). A main difference is the focal points seen in the dipole plot (blue point) are not apparent in the principal stress plot. The red dots denote the points where both principal stresses have the same value, and therefore the orientation is ill-defined. The difference between the location of the blue and the red dots shows that one cannot find the focal point using the background stress alone. }  
\label{fig:principal_stress}
\end{figure*}
\clearpage

\section*{Movie S1.}
Different crack trajectories around a flattened dislocation ($R_\mathrm{in}/R_\mathrm{out}=0.265$, $R_\mathrm{in}/b=1$). Each crack takes typically one hour to fully propagate. Cracks propagate inwards from the outer boundary only from the side to the right of the Burgers vector and not from the left. More strikingly, all propagating cracks converge towards the same focal point, located at $\theta = 0$, near, but not at $R_\mathrm{in}$.

\bibliographystyle{Science}